\documentclass[aps,prl,oneside,twocolumn]{revtex4-1}
\usepackage{amsmath}
\usepackage{amssymb}
\usepackage{graphicx}

\begin{document}
\draft \title{A quantum mechanical model of the upper bounds of the cascading contribution to the second hyperpolarizability}
\author{Nathan J. Dawson$^{\ddag,\dag,1}$, Benjamin R. Anderson$^{\ddag}$, Jennifer L. Schei$^{\ddag}$, and Mark G. Kuzyk$^{\ddag,2}$}
\address{$^{\ddag}$Department of Physics and Astronomy, Washington State University,
Pullman, WA 99164-2814 \\ $^{\dag}$ Currently at Department of Physics and Astronomy, Youngstown State University, Youngstown, OH 44555}
\email{$^1$dawsphys@hotmail.com, $^2$kuz@wsu.edu}
\date{\today}

\begin{abstract}Microscopic cascading of second-order nonlinearities between two molecules has been proposed to yield an enhanced third-order molecular nonlinear-optical response.  In this contribution, we investigate the two-molecule cascaded second hyperpolarizability and show that it will never exceed the fundamental limit of a single molecule with the same number of electrons as the two-molecule system. We show the apparent divergence behavior of the cascading contribution to the second hyperpolarizability vanishes when properly taking into account the intermolecular interactions. Although cascading can never lead to a larger nonlinear-optical response than a single molecule, it provides alternative molecular design configurations for creating materials with large third-order susceptibilities that may be difficult to design into a single molecule.
\end{abstract}

\maketitle

\section{Introduction}
\label{sec:introcasc}

Cascading is a process where photons generated by one molecule interact with nearby molecules.  A linear local field model is the simplest form of cascading.  When photons are generated through nonlinear processes that mediate interactions between molecules, more complex behavior is observed.  The net effect is two molecules that interact through fields generated by a second-order nonlinear-optical process that act as a third-order nonlinear-optical process.  Thus, it is natural to ask whether or not it is possible to make two-molecule cascaded systems with a larger nonlinear-optical response than that of a single molecule.\cite{mered82.01,mered82.04,dolga07.01,baev10.01}

There are two limiting geometries of aligned one-dimensional molecules that encompass all possible cascading susceptibilities. Previously, we discussed the cascading contribution of two side-by-side molecules with the long axis oriented in the direction of the electric field.\cite{dawson11.01a} In this paper we discuss the effects of cascading in a two-molecule system where the molecules are in an end-to-end configuration and aligned parallel to the applied electric field.

Over the last decade, it has been shown that there are fundamental limits to the nonlinear molecular susceptibilities.\cite{kuzyk00.01,kuzyk00.02,Kuzyk03.05,kuzyk03.02,kuzyk05.02,kuzyk06.03,perez08.01} These limits have yet to be broken. Moreover, experiments show a large gap between current measurements of the best second hyperpolarizabilities and the theoretical limit.\cite{Tripa04.01,zhou06.01,zhou07.02} Thus, it would be interesting if cascading could be used as a way to exceed the fundamental limits.  Since the fundamental limits hold for any quantum system, a two-molecule cascaded system should also obey the limits.  However, since the intermediate photons can, in principle, be resonantly enhanced, it is not clear whether or not the limits can indeed be broken.  This work seeks to better understand the nature of the cascading contribution to the second hyperpolarizability and to understand if it is possible to exceed the fundamental limit.


\section{Classical Route to Cascading}
\label{sec:classicalcasc}

Consider two freely-rotating one-dimensional molecules that are located on the same electric field line of a one-dimensional applied field, $\mathbf{E}_a = E_a \hat{z}$, where we are using the cartesian coordinates $x$, $y$, and $z$. When the molecules are also aligned with the field, the system is said to be in the end-to-end configuration. Since the molecules are assumed to be identical, $\left|\mathbf{p}_i\right| = \left|\mathbf{p}_j\right|$, where $\left|\mathbf{p}\right|$ represents the magnitude of the electric dipole moment.

The induced dipole moment of the $i$th molecule in terms of the applied field and the electric field of the $j$th molecule is
\begin{eqnarray}
\mathbf{p}_{i} &=& \hat{f}_i \left(\phi_i,\theta_i\right) \alpha \left[\hat{f}_i\left(\phi_i,\theta_i\right)\cdot \left(\mathbf{E}_a + \mathbf{E}_{j}\right) \right] \nonumber \\
&+& \hat{f}_i\left(\phi_i,\theta_i\right) \beta \left[ \hat{f}_i \left(\phi_i,\theta_i\right) \cdot \left(\mathbf{E}_a + \mathbf{E}_{j}\right)\right]^2 \nonumber \\
&+& \hat{f}_i\left(\phi_i,\theta_i\right) \gamma \left[ \hat{f}_i \left(\phi_i,\theta_i\right) \cdot \left(\mathbf{E}_a + \mathbf{E}_{j}\right)\right]^3 .
\label{eq:molsus3rdorder}
\end{eqnarray}
Here,
\begin{equation}
\hat{f}_i \left(\phi_i,\theta_i\right) = \sin\phi_i \sin\theta_i \hat{x}-\cos\phi_i\sin\theta_i \hat{y} + \cos\theta_i \hat{z} ,
\label{eq:fequation}
\end{equation}
where $\phi$ is the azimuthal angle and $\theta$ is the polar angle of molecule $i$. Note that $i\neq j$.  Also, $\alpha$, $\beta$, and $\gamma$ are the polarizability, hyperpolarizability, and second hyperpolarizability.  The induced dipole field of the $j$th molecule at the $i$th molecule's position in Equation \ref{eq:molsus3rdorder} is
\begin{equation}
\mathbf{E}_{j} = \frac{p_j}{r^3}\left(-\hat{x}\sin\phi_j \sin\theta_j + \hat{y}\cos\phi_j \sin\theta_j + 2\hat{z}\cos\theta_j \right) .
\label{eq:electrdiplfield}
\end{equation}

We wish to solve the self-consistent equation for $p_i\left(\theta_i,\theta_j,\phi_{ij}\right)$, where $\phi_{ij} = \phi_i-\phi_j$. The contribution of the $i$th molecule to the first three effective molecular susceptibilities, $\alpha_{\mathrm{eff},i}$, $\beta_{\mathrm{eff},i}$, and $\gamma_{\mathrm{eff},i}$, are shown in the appendix, where they are derived by solving Equation \ref{eq:molsus3rdorder} for $p_i$ and using
\begin{equation}
\xi^{\left(n\right)} = \left. \frac{1}{n!} \frac{d^n p}{d E^n} \right|_{E=0} , \label{eq:kn}
\end{equation}
where $\xi^{\left(1\right)} = \alpha$, $\xi^{\left(2\right)} = \beta$, and $\xi^{\left(3\right)} = \gamma$.

\begin{figure}[b!]
\begin{center}\includegraphics[scale=1]{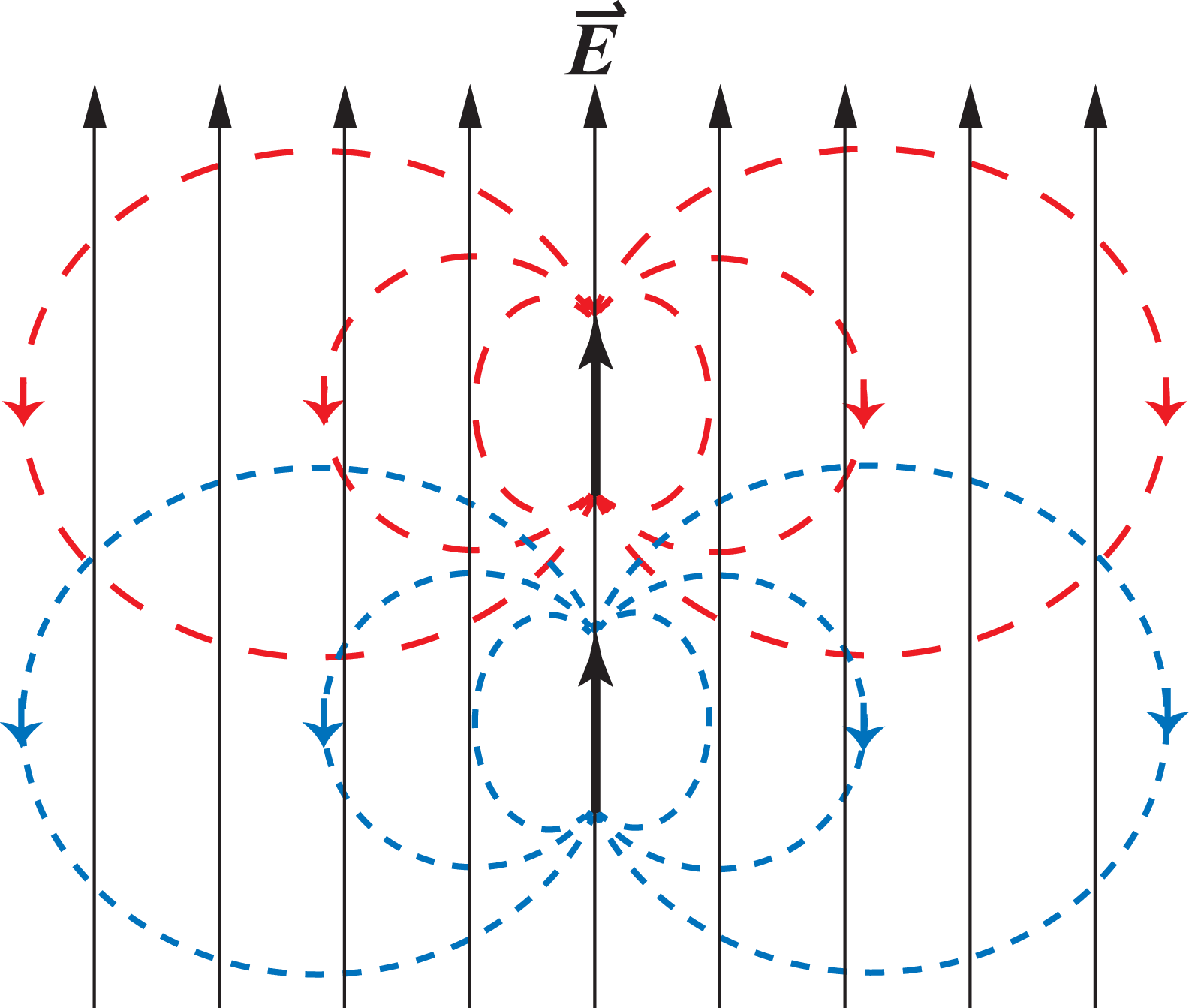}\end{center}
\caption{(Color online) The configuration of the molecules studied in this work. The two short black arrows represent the molecules while the long black arrows represent the applied electric field. The red (long dash) and blue (short dash) curves represent the electric field lines from each induced dipole.}
\label{fig:aligned1D}
\end{figure}

The polarization is maximum when the two molecules are aligned in the direction of the applied external electric field as shown in Figure \ref{fig:aligned1D}. Thus, to study the upper bound of this case, we set $\theta_1 = \theta_2 = 0$, and
\begin{equation}
E_i = E_j = \frac{2p}{r^3} .
\label{eq:alignedEp}
\end{equation}
For the perfectly aligned case, the first three effective molecular susceptibilities, $\alpha_\mathrm{eff}$, $\beta_\mathrm{eff}$, and $\gamma_\mathrm{eff}$, yield
\begin{eqnarray}
\alpha_\mathrm{eff} = 2r^3\frac{\alpha}{r^3-2\alpha} \label{eq:align1Dalpha} , \\
\beta_\mathrm{eff} = 2r^9\frac{\beta}{\left(r^3-2\alpha\right)^3} \label{eq:align1Dbeta} ,
\end{eqnarray}
and
\begin{equation}
\gamma_\mathrm{eff} = 2r^{12}\frac{\gamma\left(r^3-2\alpha\right)+4\beta^2}{\left(r^3-2\alpha\right)^5} .
\label{eq:align1Dgamma}
\end{equation}

Note that the molecular susceptibilities, Equations \ref{eq:align1Dalpha}-\ref{eq:align1Dgamma}, are in scalar form.  For typical $\pi$-conjugated donor-acceptor molecules, $\beta_{xxx} \equiv \beta$ dominates the response. All further calculations presented in this paper neglect the permanent molecular dipole moment and therefore the ground state transition moment is $x_{00} = 0$. Also, we will only consider the far off-resonant case, $E\left(\omega=0\right)$, where $\omega$ is the angular frequency.

The polarizability has units of volume, and is loosely related to the `size' of the molecule.  The self-consistent calculation assumes two distinct molecules, so their separation must be larger than their size.  Therefore, the separation must obey $r>\sqrt[3]{2\alpha}$. Moreover, the factor, $r^3/\left(r^3-2\alpha\right)$, which can be viewed as a local field factor due to interaction, can be expanded to yield
\begin{equation}
\frac{r^3}{r^3-2\alpha} = 1 + 2\frac{\alpha}{r^3} + 4\frac{\alpha^2}{r^6} + \cdots .
\label{eq:lffapprox}
\end{equation}
The first term on the righthand side of Equation \ref{eq:lffapprox} describes the classical cascading approximation with no local field, i.e. $r^3 \gg 2\alpha$. In this approximation Equation \ref{eq:align1Dgamma} reduces to
\begin{equation}
\gamma_\mathrm{eff}^{\left(0\right)} \approx 2\gamma + 8\frac{\beta^2}{r^3} ,
\label{eq:align1Dgammaapprox}
\end{equation}
where the second term on the righthand side is the cascading contribution. In the end-to-end geometry, the resultant cascading term is a factor of $8$ larger than that given by Baev \textit{et al}.\cite{baev10.01}

Before proceeding, we note that nonlinear optics is based on the series expansion. For convergence, each term must be smaller than the preceding one. When $r^3 = 2 \alpha$, the series does not converge, even for an infinitesimal applied field, since $\beta_{\mathrm{eff}}/\alpha_{\mathrm{eff}}$, $\gamma_{\mathrm{eff}}/\beta_{\mathrm{eff}}$, etc. diverge. Thus, the fundamental basis for nonlinear optics fails when $r^3\rightarrow 2^{+} \alpha$. In this limit, the system is better treated as a single molecule, in which case the fundamental limits must apply directly.

The approach of this work is to test the regime of strong molecular interactions. If the fundamental limits are not exceeded in this regime, and are known not to be exceeded for two noninteracting molecules as well as for the single merged molecule, then this would prove that molecular interactions cannot lead to a second hyperpolarizability that exceeds the limits.

In the classically interacting approximation ($r^3 \gg 2 \alpha$), Equations \ref{eq:align1Dalpha}-\ref{eq:align1Dbeta} reduce to a polarizability and hyperpolarizability of $2\alpha$ and $2\beta$, as expected for two non-interacting molecules. However, there is an additional contribution to the second hyperpolarizability due to cascading as quantified by Equation \ref{eq:align1Dgammaapprox}.

\section{Molecular Interactions}
\label{sec:interactcasc}

The above classical model assumes that the molecules are unchanged when they are in close proximity to each other.  However, we would expect interactions to cause the energy levels to shift and the transition dipole moments to be altered.  Since these effects get stronger as the molecular density increases, as do cascading effects, they need to be taken into account.

We begin by considering the energy shifts due to molecular interactions.  A system with $n$ molecules can be approximated by $M^n$ states, where $M$ is the number of states used to characterize a single molecule.  For simplicity, we will approximate each molecule using a two-state model.  The two-molecule system will then be described by four states.  Thus, the resulting two-molecule system should yield a good approximation to the nonlinear response when it approaches the fundamental limit,\cite{kuzyk00.01} in which case $E_{20}\rightarrow \infty$, where $E_{ij} = E_i - E_j$.  While real systems are known to have an infinite number of states, numerical optimization studies of potential energy functions show that a quantum system, when at the limit, is dominated by three states.\cite{zhou06.01,zhou07.01} Given that our investigations focus on these limiting cases, the three-level model is appropriate. The fourth state provides a correction factor when the hyperpolarizabilities are further from the limit.

Two molecules in close proximity experience a van der Waals interaction, which in the end-to-end geometry is given by
\begin{equation}
V\left(r\right) = \frac{2e^2x^A x^B}{r^3} ,
\label{eq:vanderwaalspert}
\end{equation}
where $r$ is the distance between the two molecules, $x^i$ is the position operator for the $i$th molecule, and $e$ is the charge of an electron in gaussian units. We use perturbation theory to describe this interaction.

For a molecule with no permanent dipole moment, the perturbed energies for the geometric configuration shown in Figure \ref{fig:aligned1D} are
\begin{eqnarray}
& & E^{S}_0 = E_0 + E_0 + \Delta E_{0,0} , \label{eq:pertE0} \\
& & E^{S}_1 = E_1 + E_0 + \Delta E_{1,0}^{-} , \label{eq:pertE1} \\
& & E^{S}_2 = E_1 + E_0 + \Delta E_{1,0}^{+} , \label{eq:pertE2} \\
& & E^{S}_3 = E_1 + E_1 + \Delta E_{1,1} ,
\label{eq:pertE3}
\end{eqnarray}
where
\begin{eqnarray}
& & \Delta E_{0,0} = -2\frac{e^4}{d^6}\frac{\left|x_{10}\right|^4}{E_{10}} , \label{eq:Echange00}\\
& & \Delta E_{1,0}^{-} = -2\frac{e^2}{d^3}\left|x_{10}\right|^2 , \label{eq:Echange10minus}\\
& & \Delta E_{1,0}^{+} = 2\frac{e^2}{d^3}\left|x_{10}\right|^2 , \label{eq:Echange10plus}\\
& & \Delta E_{1,1} = 2\frac{e^2}{d^3}\left|x_{11}\right|^2 + 2\frac{e^4}{d^6}\frac{\left|x_{10}\right|^4}{E_{10}} ,
\label{eq:Echange11}
\end{eqnarray}
and the superscript $S$  is a label for the coupled system. The transition moments $x_{ij}$ and energies $E_j$ are for the unperturbed molecules.  For a molecule with no permanent dipole moment, $x_{00} = 0$, $x_{11}$ can be replaced with $\Delta x_{10} = x_{11}-x_{00}$ in the first term on the righthand side of Equation \ref{eq:Echange11} - allowing the sum rules to be used to recast the above expressions.

Next, we consider the effect of dipole interactions on the transition moments.  The diagonalized wavefunctions for the four-level system are
\begin{eqnarray}\label{eq:wavefunction1}
& & \left|0^S\right\rangle = \left|0_A\right\rangle \left|0_B\right\rangle , \\
& & \left|1^S\right\rangle = \frac{1}{\sqrt{2}}\left(\left|0_A\right\rangle \left|1_B\right\rangle - \left|1_A\right\rangle \left|0_B\right\rangle\right) , \\
& & \left|2^S\right\rangle = \frac{1}{\sqrt{2}}\left(\left|0_A\right\rangle \left|1_B\right\rangle + \left|1_A\right\rangle \left|0_B\right\rangle\right) , \\
& & \left|3^S\right\rangle = \left|1_A \right\rangle\left|1_B\right\rangle, \label{eq:wavefunction2}
\end{eqnarray}
where the subset of degenerate states has been diagonalized with respect to the interaction Hamiltonian. Since there is no applied magnetic field, and thus no magnetic vector potential in the conservative Hamiltonian, the wavefunctions can be chosen to yield real transition moments. As a result, the transition moment matrix is symmetric, that is, $x_{ij} = x_{ji}$. Furthermore, the molecules are indistinguishable, $x^{A}_{ij} = x^{B}_{ij}$.

\begin{table}[b!]
\caption{Transition moments} 
\centering
\begin{tabular}{c c c} 
\hline\hline 
Transition Moment & Molecule $A$ & Molecule $B$ \\ [0.5ex] 
\hline
$\left\langle 0^S\right|x^i \left|0^S\right\rangle$ & $0$ & $0$\\
$\left\langle 0^S\right|x^i \left|1^S\right\rangle$ & $- \frac{1}{\sqrt{2}}x_{01}^A$ & $\frac{1}{\sqrt{2}}x_{01}^B$\\
$\left\langle 0^S\right|x^i \left|2^S\right\rangle$ & $\frac{1}{\sqrt{2}}x_{01}^A$ & $\frac{1}{\sqrt{2}}x_{01}^B$\\
$\left\langle 0^S\right|x^i \left|3^S\right\rangle$ & $0$ & $0$\\
$\left\langle 1^S\right|\bar{x}^i \left|1^S\right\rangle$ & $\frac{1}{2}\Delta x_{10}^A$  & $\frac{1}{2}\Delta x_{10}^B$\\
$\left\langle 1^S\right|\bar{x}^i \left|2^S\right\rangle$ & $-\frac{1}{2}\Delta x_{10}^A$  & $\frac{1}{2}\Delta x_{10}^B$\\
$\left\langle 1^S\right|\bar{x}^i \left|3^S\right\rangle$ & $\frac{1}{\sqrt{2}}x_{01}^A$ & $-\frac{1}{\sqrt{2}}x_{01}^B$\\
$\left\langle 2^S\right|\bar{x}^i \left|2^S\right\rangle$ & $\frac{1}{2}\Delta x_{10}^A$  & $\frac{1}{2}\Delta x_{10}^B$\\
$\left\langle 2^S\right|\bar{x}^i \left|3^S\right\rangle$ & $\frac{1}{\sqrt{2}}x_{01}^A$ & $\frac{1}{\sqrt{2}}x_{01}^B$\\
$\left\langle 3^S\right|\bar{x}^i \left|3^S\right\rangle$ & $\Delta x_{10}^A$ & $\Delta x_{10}^B$\\
\hline
\end{tabular}
\label{table:systemtransmoments}
\end{table}

The linear and nonlinear susceptibilities are calculated using a sum-over states expression.  For the two-molecule system, the transition moments are calculated using the wavefunctions given by Equations \ref{eq:wavefunction1}-\ref{eq:wavefunction2}, the position operator $x = x^A + x^B$, and the energies given by Equations \ref{eq:pertE0}-\ref{eq:pertE3}.  Using this model of the van der Waals interaction, the second hyperpolarizability is a simple sum of the form $\gamma^S = \gamma^A + \gamma^B$, and the direct second hyperpolarizability contributed by molecule $i$ is given by
\begin{eqnarray}
\gamma^i &=& 4 e^4 \displaystyle\sum_{n,m,l}^3 \hspace{-.2cm}\left.\right.^\prime \frac{\left\langle 0^S\right|x^i\left|n^S\right\rangle \left\langle n^S\right|\bar{x}^i\left|m^S\right\rangle }{E_{n0}^S E_{m0}^S E_{l0}^S} \nonumber \\
&\times& \left\langle m^S\right|\bar{x}^i\left|l^S\right\rangle \left\langle l^S\right|x^i\left|0^S\right\rangle \nonumber \\
&-& 4 e^4 \displaystyle\sum_{n,m}^3 \hspace{-.1cm}\left.\right.^\prime \frac{\left\langle 0^S\right|x^i\left|n^S\right\rangle \left\langle n^S\right|x^i\left|0^S\right\rangle }{\left(E_{n0}^S\right)^2 E_{m0}^S} \nonumber \\
&\times& \left\langle 0^S\right|x^i\left|m^S\right\rangle \left\langle m^S\right|x^i\left|0^S\right\rangle ,
\label{eq:gammasumfull}
\end{eqnarray}
where the prime denotes that the sum excludes the ground state.  It is straightforward to show that there are no cross terms of the form $x^A x^B$.  Likewise, the hyperpolarizability contributed by each molecule is
\begin{eqnarray}
&\beta^i& = -3e^3 \times \nonumber \\
& & \displaystyle\sum_{n,m}^3 \hspace{-.08cm}\left.\right.^\prime \frac{\left\langle 0^S\right|x^i\left|n^S\right\rangle \left\langle n^S\right|\bar{x}^i\left|m^S\right\rangle \left\langle m^S\right|x^i\left|0^S\right\rangle}{E_{n0}^S E_{m0}^S} ,
\label{eq:betasumfull}
\end{eqnarray}
where $\beta^S = \beta^A + \beta^B$.

The barred position operator is defined as $\left\langle i\right|\bar{x}\left|j\right\rangle \equiv \left\langle i\right|x\left|j\right\rangle - \left\langle 0\right|x\left|0\right\rangle\delta_{ij}$, where $\delta_{ij}$ is the Kronecker delta.\cite{orr71.01} Thus,
\begin{eqnarray}
\left\langle i\right|\bar{x}^A \left|j\right\rangle &=& \left\langle k_A\right| \left\langle l_B\right| \bar{x}^A \left|m_A\right\rangle \left|n_B\right\rangle \nonumber \\
&=& \left(\left\langle k_A\right| x^A \left|m_A\right\rangle - \left\langle 0_A\right| x^A \left|0_A\right\rangle \delta_{km} \right) \delta_{ln} . \nonumber \\
\label{eq:coupledbarx}
\end{eqnarray}
The transition dipole moments for molecule $A$ and $B$ are shown in Table \ref{table:systemtransmoments}.

The transition energies used to calculate $\beta$ and $\gamma$ from Equations \ref{eq:gammasumfull} through \ref{eq:betasumfull} are derived from Equations \ref{eq:pertE0}-\ref{eq:Echange11} and are of the form,
\begin{eqnarray}
E_{10}^S = E_{10} - \frac{2e^2}{r^3}\left|x_{01}\right|^2 + \frac{2e^4}{r^6}\left|x_{01}\right|^4 , \label{eq:transenergy10} \\
E_{20}^S = E_{10} + \frac{2e^2}{r^3}\left|x_{01}\right|^2 + \frac{2e^4}{r^6}\left|x_{01}\right|^4 , \label{eq:transenergy20} \\
E_{30}^S = 2E_{10} + \frac{2e^2}{r^3}\left|\Delta x_{10}\right|^2 + \frac{4e^4}{r^6}\left|x_{01}\right|^4 . \label{eq:transenergy30}
\end{eqnarray}

\section{Cascading and the fundamental limit}
\label{sec:cascfundlimit}

\subsection{Case 1: Quantum and classical interactions}
\label{sub:interactingcase}

The generalized Thomas-Kuhn sum rules have been used to calculate the upper limit for the nonlinear molecular susceptibilities,\cite{thoma25.01,kuzyk00.01} and are calculated without approximation using the double commutation relationship between a conservative Hamiltonian and the position operator. The resulting sum rules are of the form,
\begin{equation}
\displaystyle\sum_{n=0}^\infty \left(E_n-\frac{1}{2}\left(E_q + E_p\right)\right)x_{qn}x_{np} = \frac{\hbar^2 N}{2 m}\delta_{q,p} ,
\label{eq:sumrules}
\end{equation}
and relate the matrix elements of the position operator to the energies. All solutions to the Schrodinger equation must obey the sum rules.

The three-level ansatz states that when a quantum system has a nonlinear susceptibility that is near the fundamental limit, three states contribute.\cite{zhou06.01,zhou08.01,kuzyk09.01} Thus, the four-level system resulting from two interacting molecules should be sufficient for studying the behavior of the largest cascaded nonlinear response. In the three-level ansatz, the normalized transition moment from the ground state to first excited state, $X$, is an important parameter, which is given by
\begin{equation}
X = \frac{\left|x_{10}\right|}{\left|x_{10}^\mathrm{max}\right|} ,
\label{eq:xfraction}
\end{equation}
where $x_{10}^\mathrm{max}$ is determined from the sum rules to be of the form
\begin{equation}
\left|x_{10}^\mathrm{max}\right|^2 = \frac{\hbar^2 N}{2 m E_{10}} .
\label{eq:x10max}
\end{equation}

Setting $q=0$, we can multiply both sides of Equation \ref{eq:sumrules} by $x_{p0}$ and sum over $p$, which gives
\begin{equation}
\left|x_{n0}\right|^2 \Delta x_{n0} = - \displaystyle\sum_{p \neq n}^\infty \frac{E_{pn}+E_{p0}}{E_{n0}}x_{0n}x_{np}x_{p0} .
\label{eq:diagsolver}
\end{equation}
Under the three-level ansatz, one can use the sum rules to find $x_{12}$. Then, using Equation \ref{eq:diagsolver} and allowing $E_{20}\rightarrow\infty$, we get
\begin{equation}
\left|x_{10}\right| \Delta x_{10} = -2\left|x_{10}^\mathrm{max}\right|^2\sqrt{1-X^4} .
\label{eq:deltax10}
\end{equation}

Regardless of the opposite signs for some of the entries in Table \ref{table:systemtransmoments}, $\beta^A = \beta^B$ and $\gamma^A = \gamma^B$. To simplify the analysis that follows, we define the dimensionless parameter,
\begin{equation}
r^\prime = r/\sqrt[3]{\alpha_\mathrm{max}} ,
\label{eq:rprime}
\end{equation}
where
\begin{equation}
\alpha_{\mathrm{max}} = \frac{e^2 \hbar^2 N}{m E_{10}^2} .
\label{eq:alphamaxnote}
\end{equation}
The individual molecular contributions, $\beta_\mathrm{pert}$ and $\gamma_\mathrm{pert}$, for the interacting system of two-level molecules with perturbed energy states in terms of $r^\prime$ and $X$ are calculated from Equation \ref{eq:betasumfull},
\begin{equation}
\beta_\mathrm{pert} = 6\sqrt{2}\frac{e^3 \hbar^3 N^{3/2}}{m^{3/2} E_{10}^{7/2}} \frac{r^{\prime 12} X \sqrt{1 - X^4} \left(2 r^{\prime 6} + X^4\right)^2}{\left(4r^{\prime 12}+X^8\right)^2} ,
\label{eq:beta2Lpert}
\end{equation}
and Equation \ref{eq:gammasumfull},
\begin{eqnarray}
\gamma_\mathrm{pert} &=& \frac{e^4\hbar^4N^2}{m^2E_{10}^5} \left[ \left(32r^{\prime 21}+4r^{\prime 3
}X^{12}\right) \left(4-9X^4+5X^8\right) \right. \nonumber \\
&+& \left(16r^{\prime 15}X^4+8r^{\prime 9}X^8\right) \left(12-29X^4+17X^8\right) \nonumber \\
&-& \left(8r^{\prime 18}X^6+2r^{\prime 6}X^{14}\right) \left(21X^4-16\right) \nonumber \\
&-& \left(16r^{\prime 24}X^2+X^{18}\right) \left(5X^4-4\right) \nonumber \\
&-& \left. 8r^{\prime 12}X^{10}\left(17X^4-12\right)\right] / \nonumber \\
& & \hspace{-.35cm}\left[ \left(4r^{\prime 12}+X^8\right)^3 \left(2r^{\prime 6}X^2+4r^{\prime 3}\left(1-X^4\right)+X^6\right) \right] . \nonumber \\
\label{eq:gamma2Lpert}
\end{eqnarray}

\begin{figure}[t!]
\begin{center}\includegraphics[scale=1]{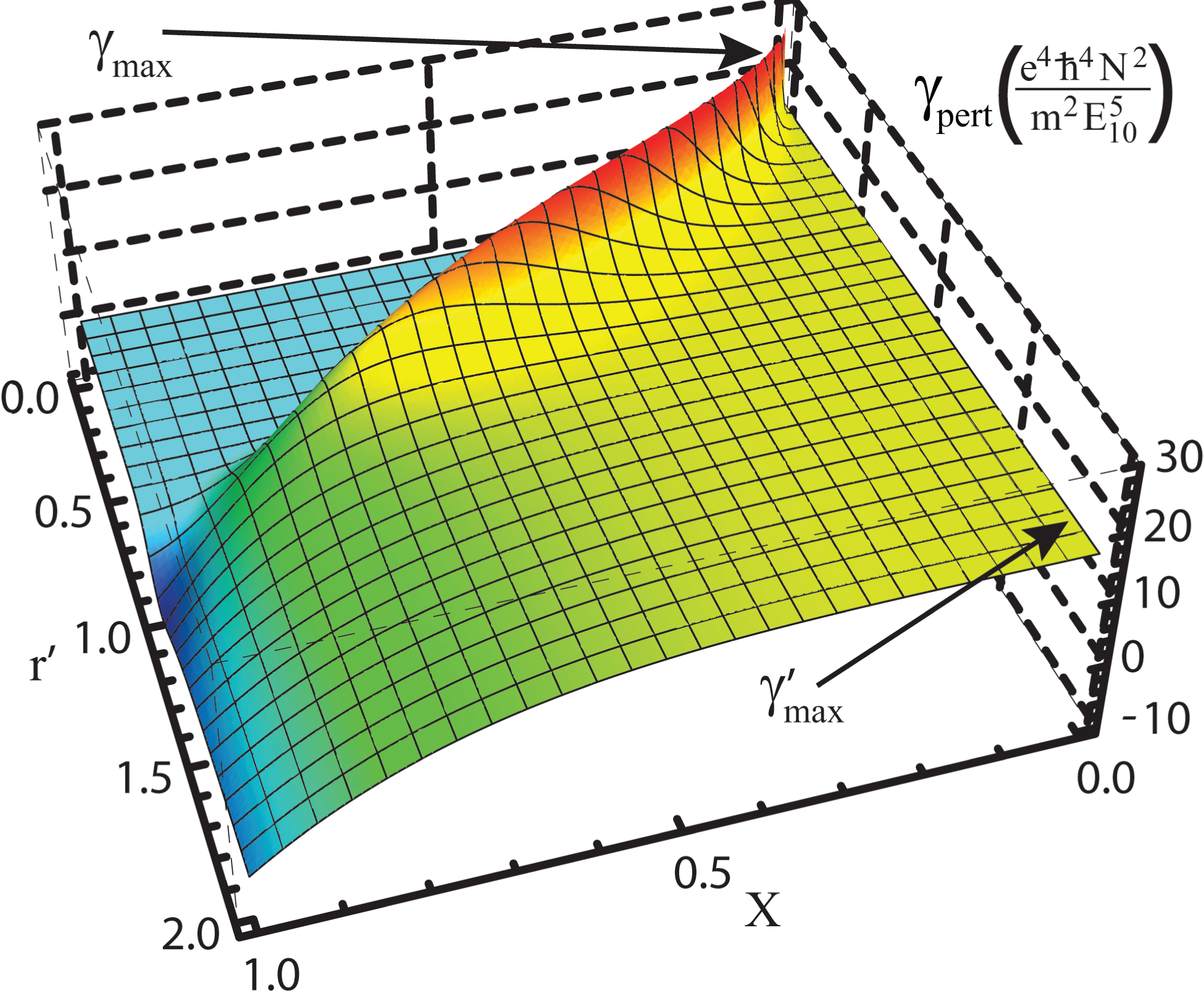}\end{center}
\caption{(Color online) The four-level quantum model of the second hyperpolarizability of two single molecules. When $r^\prime < \sqrt[3]{2}$, the two molecules overlap and the calculations break down.}
\label{fig:gmax}
\end{figure}

It should be noted that these solutions are based on perturbation calculations, which are only valid when the distance between molecules is sufficiently large to lead to small energy shifts. Since the electron cloud defines a physical size of a molecule and two molecules cannot overlap, a minimum possible separation is defined, as has been described for van der Waals forces between neighboring molecules.\cite{bk:stone02.02,bk:atkins97} There are also retardation effects which can lower the cascading contribution at large distances. However, such effects are of no concern here since we are considering the maximum cascading contribution for two molecules in close proximity.

\begin{figure}[b!]
\begin{center}\includegraphics[scale=1]{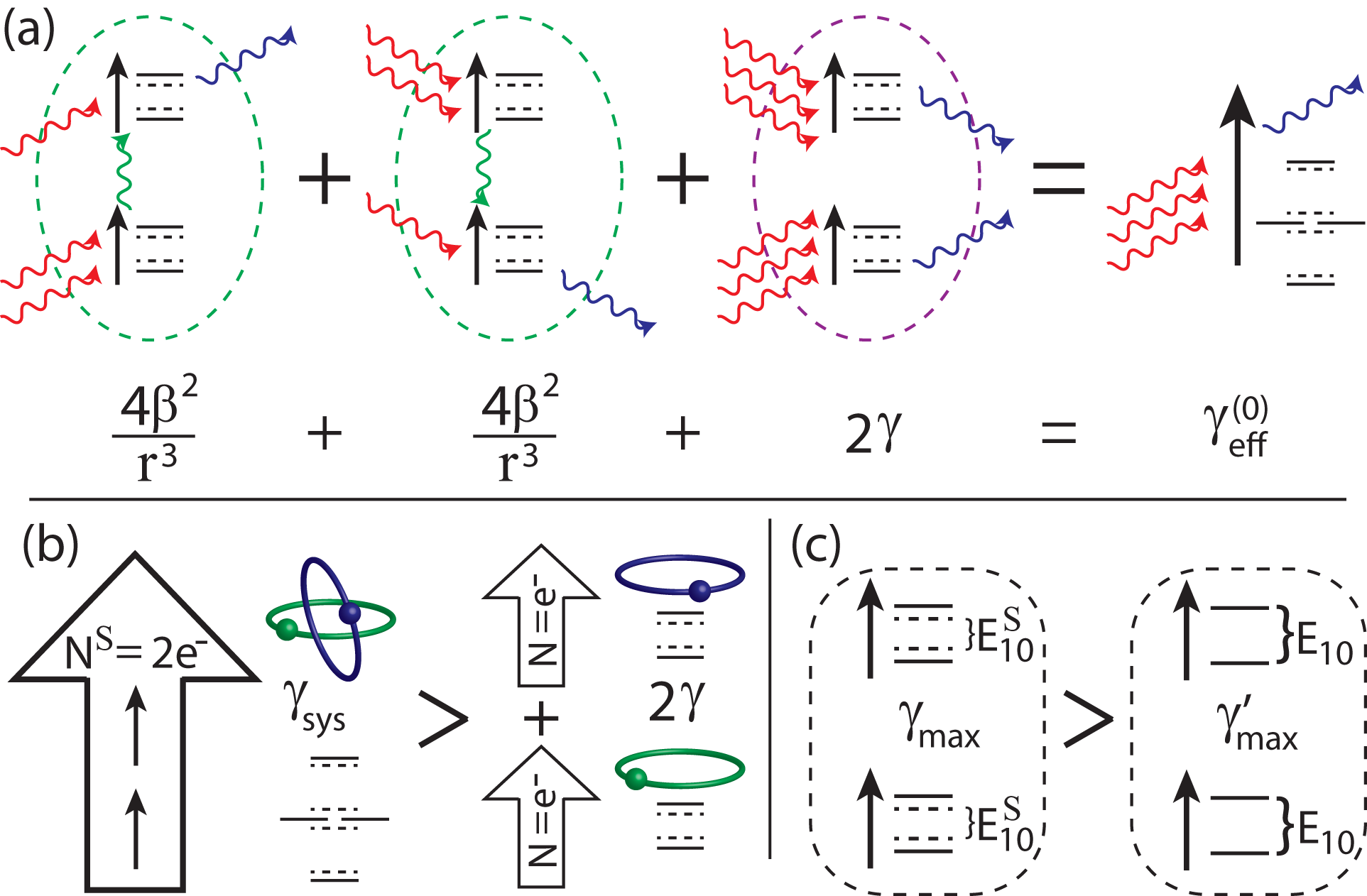}\end{center}
\caption{(Color online) A diagram illustrating the theoretical approach to cascading: (a) The relationship between $\gamma_{\mathrm{eff}}^{\left(0\right)}$ and the individual contributions, (b) the second hyperpolarizability of two molecules each with a set number of electrons with respect to the second hyperpolarizability of a single molecule with the same number of electrons, and (c) a comparison between the interacting maximum, $\gamma_{\mathrm{max}}$, and the noninteracting case, $\gamma_{\mathrm{max}}^{\prime}$.}
\label{fig:geffequation}
\end{figure}

The fundamental limit of the second hyperpolarizability is
\begin{equation}
\gamma_\mathrm{max} = 4\frac{e^4 \hbar^4 \left(N^S\right)^2}{m^2 \left(E_{10}^S\right)^5} ,
\label{eq:gammalimitsystem}
\end{equation}
where the superscript $S$ denotes the values of those parameters for the two-molecule system.  Note that the two-molecule system is characterized by a lower transition energy and has twice as many electrons as does a single molecule. The fundamental limit of the second hyperpolarizability for two molecules, each with $N$ electrons, scales as $\left(N^\mathrm{S}\right)^2 = 4N^2$, thereby quadrupling the limit relative to the single molecule.

The transition energy between the first excited state and the ground state becomes smaller as the two molecules are brought together, and therefore $E_{10}^S \leq E_{10}$ - further increasing the limit. In the semi-classical approximation, the basis of dielectric theory, the linear and nonlinear susceptibilities do not change in the presence of perturbations, thus, the eigenenergies of the underlying molecules are inviolate.  The \textit{semi-classical fundamental} limit, at $r\rightarrow\infty$ and $X=0$, can then be expressed as a function of the number of electrons in each molecule, $N$, and transition energy, $E_{10}$, of each molecule,
\begin{equation}
\gamma_\mathrm{max}^\prime \equiv 16\frac{e^4 \hbar^4 N^2}{m^2 E_{10}^5} .
\label{eq:gammalimitinequality}
\end{equation}
Thus, the semi-classical approximation yields a limit that is smaller than the exact one, or $\gamma_\mathrm{max}^\prime \leq \gamma_\mathrm{max}$. Therefore, $\gamma_\mathrm{eff}/\gamma_\mathrm{max}^\prime \geq \gamma_\mathrm{eff}/\gamma_\mathrm{max}$ and the semi-classical theory will give a larger intrinsic value than the exact result. If this ratio does not exceed unity in the semi-classical approximation, it will not do so for the exact case.

Although $\gamma_\mathrm{max}$ for specific values of $X$ and $r^\prime$ is difficult to calculate analytically because Equation \ref{eq:gamma2Lpert} is a high-degree polynomial of $X$ and $r^\prime$, it can be numerically evaluated. The numerical solution yields $\gamma_\mathrm{max} \approx 1.76 \gamma_\mathrm{max}^\prime$. However, this large value is found when $r^\prime < \sqrt[3]{2}$, the domain where the separation between the molecules is less than their size - a physically unallowable regime. Figure \ref{fig:gmax} shows a plot of $\gamma_{\mathrm{sys}} = \gamma_{\mathrm{pert}} \left(N^S,E_{10}^S\right)$ using Equation \ref{eq:gamma2Lpert} for the two molecule system, where $\gamma_{\mathrm{max}}$ and $\gamma_{\mathrm{max}}^\prime$ are derived. Note that $\gamma_{\mathrm{max}}^\prime = \gamma_{\mathrm{sys}} \left(X=0,r\rightarrow\infty\right)$. If the intrinsic effective second hyperpolarizability in the semi-classical approximation, $\gamma_\mathrm{eff}/\gamma_\mathrm{max}^\prime$, of the coupled system never exceeds unity, then $\gamma_\mathrm{eff} < \gamma_\mathrm{max}^\prime < \gamma_\mathrm{max}$. If this is the case, then the effective second hyperpolarizability of two cascaded molecules will never exceed that of one molecule with twice the number of electrons. Figure \ref{fig:geffequation} is a schematic diagram that illustrates the various terms in the cascading process and the the energy levels in each case.

\begin{figure}[b!]
\begin{center}\includegraphics[scale=1]{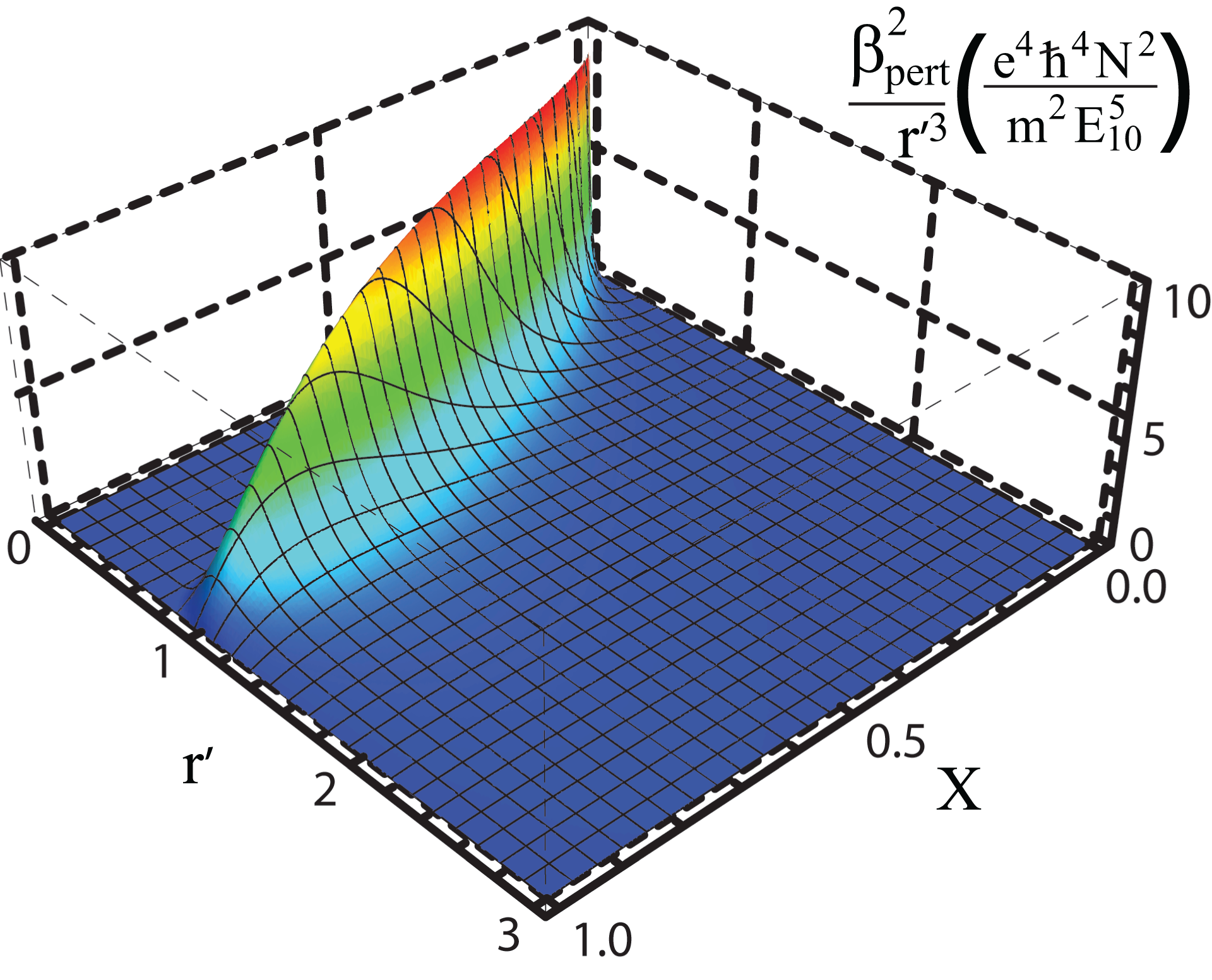}\end{center}
\caption{(Color online) The cascading contribution of $\beta_{\mathrm{pert}}^{2}/r^{\prime 3}$. The perturbed Hamiltonian keeps the function from diverging as $r^\prime \rightarrow 0$.}
\label{fig:betasqr}
\end{figure}

The effective second hyperpolarizability has a contribution from the hyperpolarizability due to the second term on the right hand side of Equation \ref{eq:align1Dgammaapprox}. Figure \ref{fig:betasqr} shows the two-level model for $\beta_{\mathrm{pert}}^{2}/r^{\prime 3}$ term in Equation \ref{eq:align1Dgammaapprox} as a function of $X$ and $r^\prime$ in units of $e^4 \hbar^4 N^2/m^2 E_{10}^5$. There is no cascading contribution at large distances, as expected.

\begin{figure}[b!]
\begin{center}\includegraphics[scale=1]{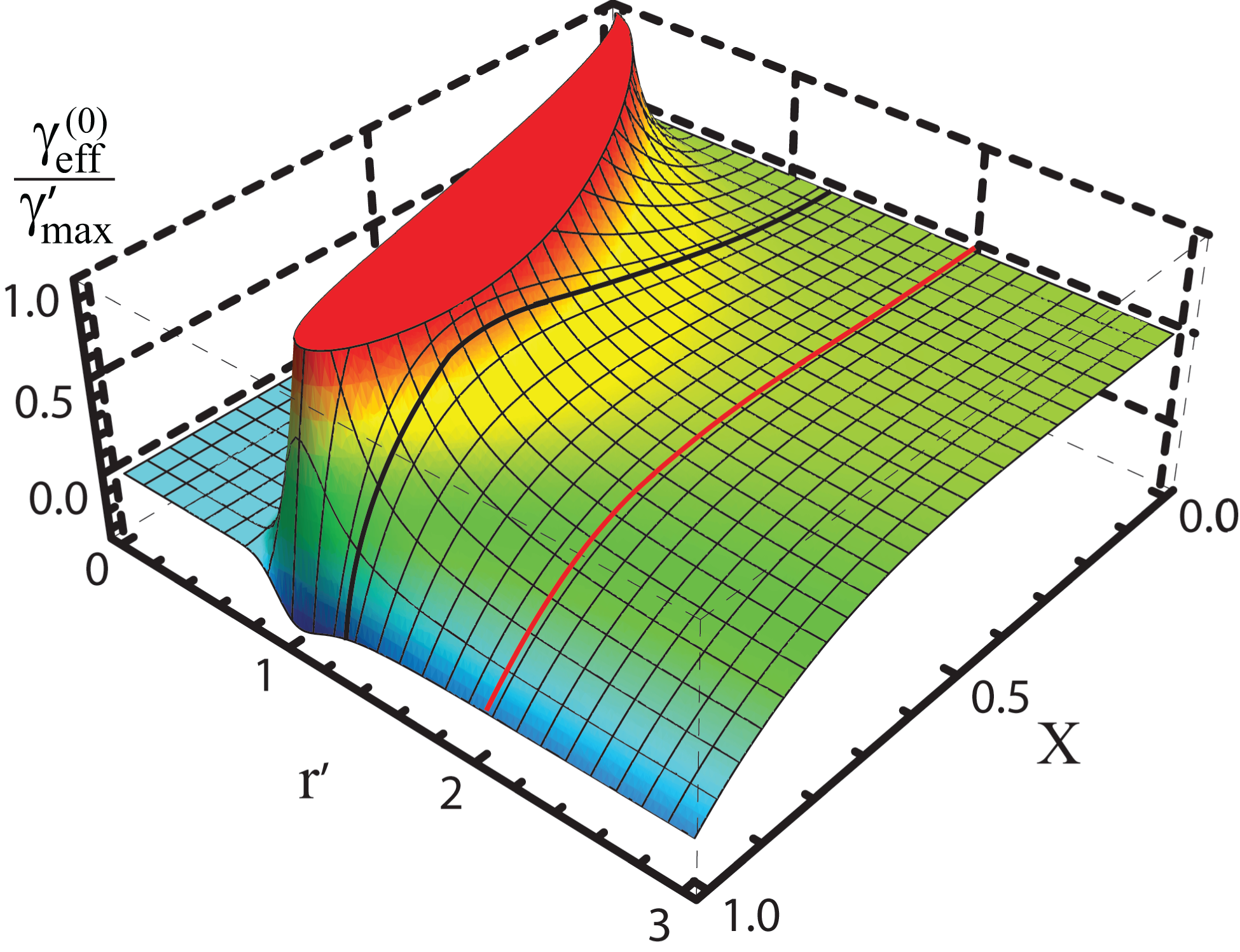}\end{center}
\caption{(Color online) The intrinsic effective second hyperpolarizability of two interacting molecules. The thick black line is the lower limit of the center of mass displacement given by the condition $r^\prime > \sqrt[3]{2}$. The thick red line is the experimental limit of closest approach for spherical atoms and molecules with large polarizabilities, $r^\prime > 2$.\cite{ghant93.01} The values that exceed the semi-classical limit have been removed from the diagram, and are shown as the red plateau.}
\label{fig:alignedcutoff}
\end{figure}

Substituting the perturbed system's values for $\gamma_{\mathrm{pert}}$ and $\beta_{\mathrm{pert}}$ from Equations \ref{eq:beta2Lpert} and \ref{eq:gamma2Lpert} into Equation \ref{eq:align1Dgammaapprox} gives the effective second hyperpolarizability with corrected transition energies. Figure \ref{fig:alignedcutoff} is a surface plot of the effective second hyperpolarizability with perturbed energies divided by $\gamma_\mathrm{max}^\prime$. As $r\rightarrow\infty$ the intrinsic hyperpolarizability converges to $1/2$, as is expected for the limit of a two electron molecule, $\gamma_\mathrm{sys}$, having $\left(N^\mathrm{S}\right)^2 = \left(2N\right)^2 = 4N^2$. At very large distances, the two molecules are independent with $\gamma = \gamma^A + \gamma^B \propto 2N^2$.   Note that the red plateau region in Figure \ref{fig:alignedcutoff} shows the domain where the limit would have been exceeded.

The second hyperpolarizability never exceeds the fundamental limit for $r^\prime\rightarrow 2^{+}$. This is an important consideration for molecular design because increasing the polarizability increases the minimum required separation distance, thereby reducing the cascading contribution. Moreover, the limit is never surpassed for any distance that is greater than $\sqrt[3]{2\alpha}$. This length is of critical important because any molecular separation that is less than this distance is unphysical and can lead to a divergence of the local field factor.

\subsection{Case 2: Only classical interactions}
\label{sub:noninteractingcase}

While a quantum characterization of interactions between two molecules leads to interesting insights, it is too complex to solve. In this section, we consider the special case where the energies are unchanged by the interaction, and we use the classical cascading approximation given by Equation \ref{eq:align1Dgammaapprox}.

Due to the simplicity of the above approximations, we choose to describe each molecule using three states. We define the fractional transition energy, $E$, as
\begin{equation}
E = \frac{E_{10}}{E_{20}} .
\label{eq:Efraction}
\end{equation}
When $E_{20} = E_{10}$, $E = 1$, and when $E_{20} \rightarrow\infty$, $E = 0$.  Thus, the system's energy spacing is fully parameterized by E in the range from zero to unity.

The sum rules for a three level model have been shown to give \cite{kuzyk00.01}
\begin{equation}
\left|x_{12}\right| = \left|x_{10}^{\mathrm{max}}\right|\sqrt{\frac{E}{1-E}} \sqrt{1+X^2} .
\label{eq:x12}
\end{equation}
Substituting Equation \ref{eq:x12} into Equation \ref{eq:diagsolver} gives
\begin{equation}
\left|x_{10}\right| \Delta x_{10} = \left|x_{10}^{\mathrm{max}}\right|^2 \frac{E-2}{\sqrt{1-E}}\sqrt{1-X^4} ,
\label{eq:deltax10threelevel}
\end{equation}
which is the three level generalization of Equation \ref{eq:deltax10}. This can be understood by letting $E_{20}\rightarrow\infty$, thereby reducing Equation \ref{eq:deltax10threelevel} to Equation \ref{eq:deltax10}. Similarly, for the second state, we get
\begin{equation}
\left|x_{20}\right| \Delta x_{20} = \left|x_{10}^{\mathrm{max}}\right|^2 \left(1-2E\right) \sqrt{\frac{E}{1-E}} X \sqrt{1+X^2} .
\label{eq:deltax20}
\end{equation}

The hyperpolarizability and the second hyperpolarizability of a single molecule with no external influences can then be written as
\begin{equation}
\beta = 3 e^3\displaystyle\sum_{n,m}^\infty \hspace{-.08cm}\left.\right. ^\prime \frac{x_{0n} \bar{x}_{nm}x_{m0}}
{E_{n0}E_{m0}}
\label{eq:betasumOW}
\end{equation}
and
\begin{eqnarray}
\gamma &=& 4 e^4\displaystyle\sum_{n,m,l}^\infty \hspace{-.16cm}\left.\right. ^\prime \frac{x_{0n}\bar{x}_{nm}
\bar{x}_{ml}x_{l0}}{E_{n0}E_{m0}E_{l0}} \nonumber \\
&-& 4 e^4 \displaystyle\sum_{n,m}^\infty \hspace{-.08cm}\left.\right. ^\prime
\frac{x_{0n}x_{n0}x_{0m}x_{m0}}{E_{n0}^2E_{m0}} .
\label{eq:gammasumOW}
\end{eqnarray}
These sum-over-states expressions were first derived using quantum perturbation methods by Orr and Ward.\cite{orr71.01}

\begin{figure}[t!]
\begin{center}\includegraphics[scale=1]{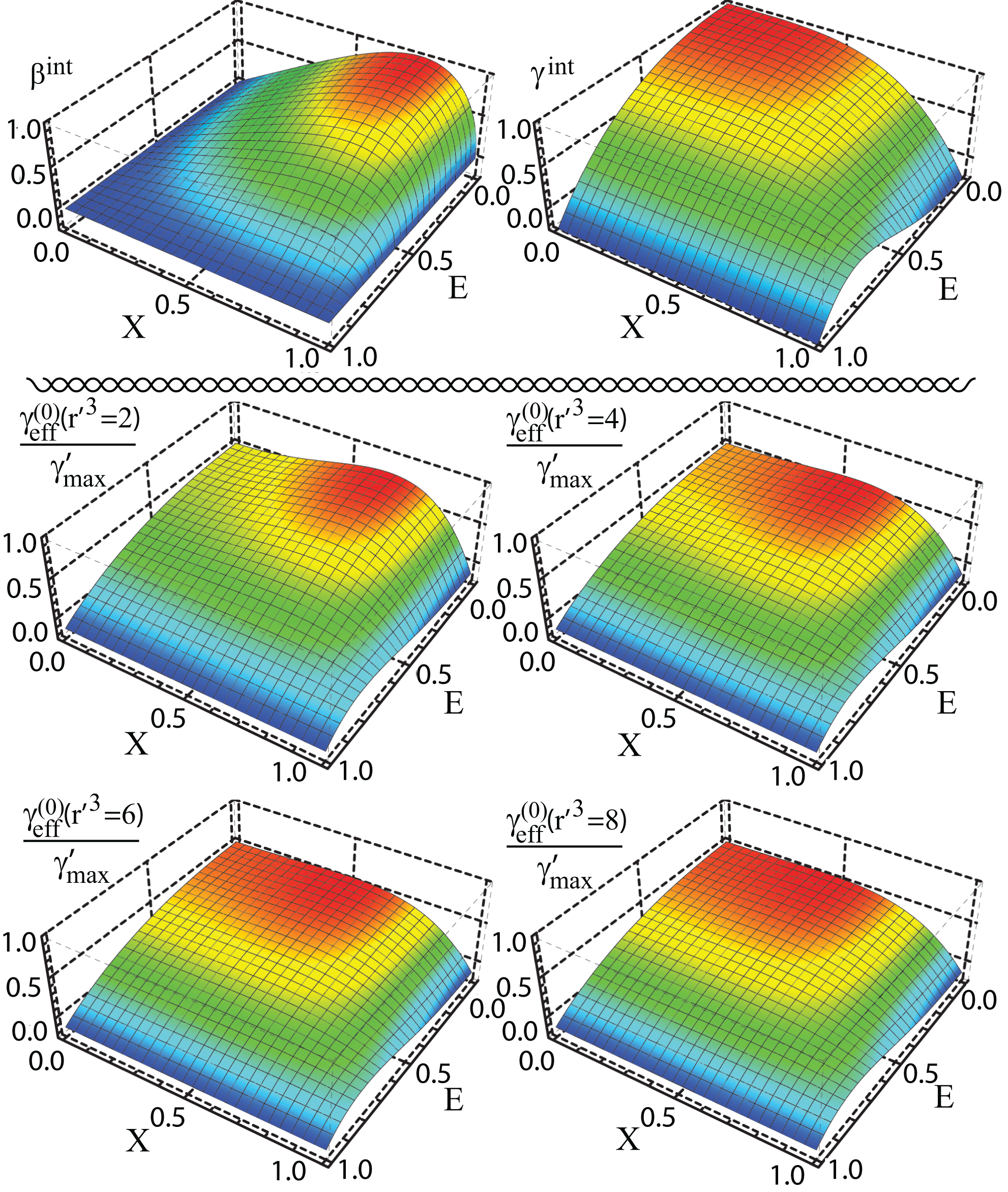}\end{center}
\caption{(Color online) The top two graphs are the intrinsic hyperpolarizability and second hyperpolarizability as a function of $X$ and $E$ for a single molecule system. The bottom four graphs show the intrinsic effective second hyperpolarizability of the aligned molecules with no energy corrections as a function of $X$ and $E$ for several $r^\prime$ values. The $\beta$ contribution decreases as the distance of separation increases.}
\label{fig:g3L}
\end{figure}

\begin{figure}[t!]
\begin{center}\includegraphics[scale=1]{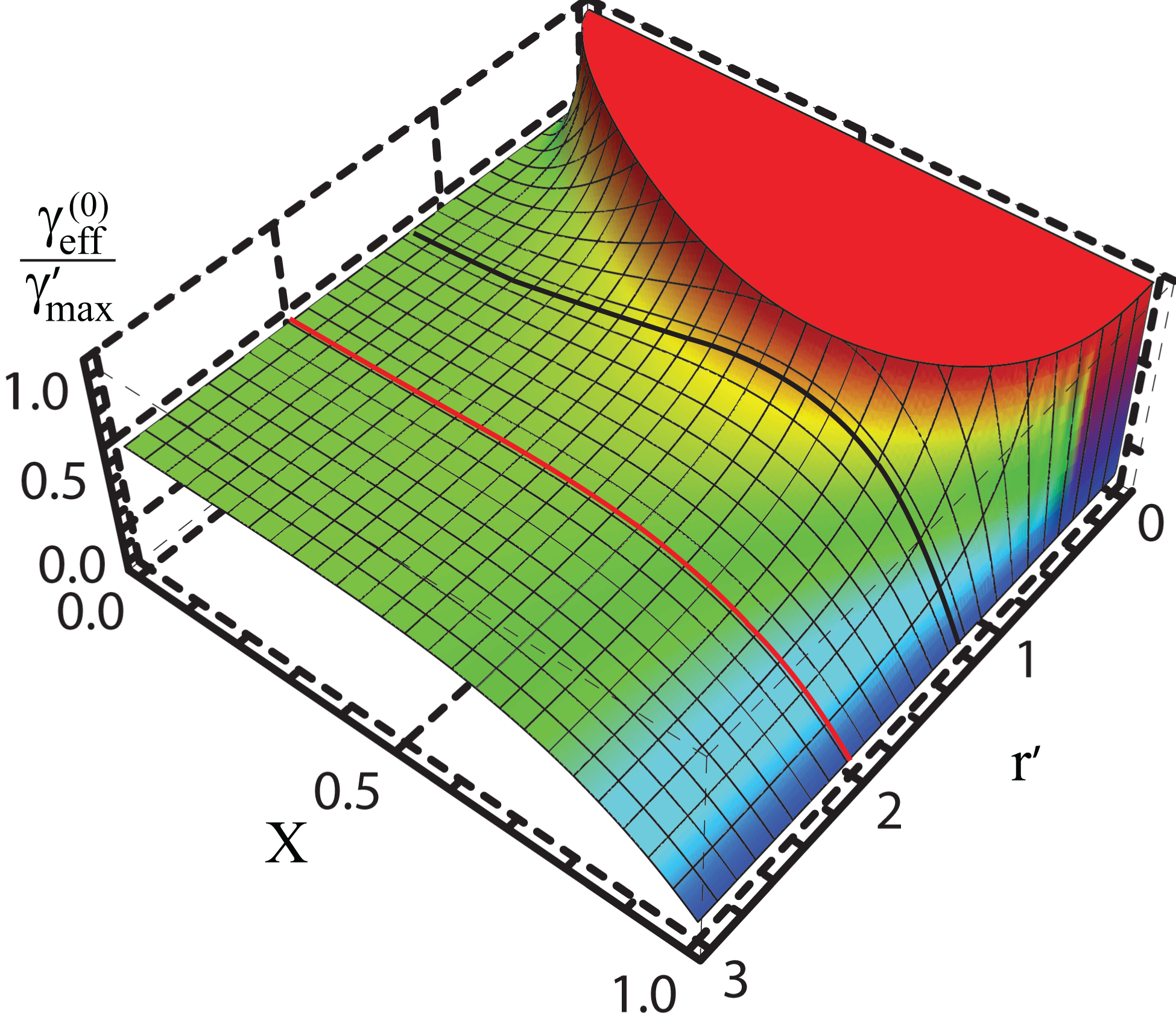}\end{center}
\caption{(Color online) The semi-classical intrinsic effective second hyperpolarizability of two end-to-end molecules. The thick black line is the minimum separation distance given by the electromagnetic size of the molecule, or $r^\prime > \sqrt[3]{2}$. The thick red line is the experimental limit of closest approach for spherical atoms which gives $r^\prime > 2$.\cite{ghant93.01} The red plateau is the region where $\gamma_{\mathrm{eff}}^{\left(0\right)}/\gamma_{\mathrm{max}}^{\prime} \leq 1$.}
\label{fig:geffunpert}
\end{figure}

Substituting Equations \ref{eq:xfraction}, \ref{eq:x10max}, and \ref{eq:Efraction}-\ref{eq:deltax20} into Equations \ref{eq:betasumOW} and \ref{eq:gammasumOW}, we find the unperturbed three level expressions for $\beta$ and $\gamma$ in terms of $X$ and $E$, which are
\begin{eqnarray}
\beta_{3L} &=& \frac {3} {\sqrt{2}} \frac{e^3 \hbar^3}{E_{10}^{7/2}}\left(\frac{N}{m}\right)^{3/2} \left(1-E\right)^{3/2} \nonumber \\ &\times& \left(1+\frac{3}{2}E+E^2\right)X\sqrt{1-X^4}
\label{eq:beta3L}
\end{eqnarray}
and
\begin{eqnarray}
\gamma_{3L} &=& \frac{e^4 \hbar^4 N^2}{m^2 E_{10}^5} \left[4 - 2(E^2-1)E^3 X^2 \right. \nonumber \\
&-& 5 \left(E-1\right)^2\left(E+1\right)\left(E^2+E+1\right)X^4 \nonumber \\
&-& \left. \left(E^3+E+3\right)E^2 \right].
\label{eq:gamma3L}
\end{eqnarray}

The classically interacting effective second hyperpolarizability is found by substituting Equations \ref{eq:beta3L} and \ref{eq:gamma3L} into Equation \ref{eq:align1Dgammaapprox}. Figure \ref{fig:g3L} shows the intrinsic effective hyperpolarizability for various separation distances. The second hyperpolarizability at the minimum separation distance, $r^\prime = \sqrt[3]{2}$, never exceeds the fundamental limit. For a single molecule, the intrinsic second hyperpolarizability is bounded by $-1/4 \leq \gamma^{\mathrm{int}} \leq 1$.

To illustrate how the effective second hyperpolarizability behaves as a function of $r^\prime$, we set $E=0$ so that each molecule is described by a two-level model that obeys the sum rules. The off-resonant nonlinear molecular susceptibilities are maximized in the two-level limit as shown in the top two graphs of Figure \ref{fig:g3L}. This case models the effective second hyperpolarizability as a function of the separation distance for two interacting molecules with large nonlinearities.

In the absence of a second excited state, the expressions given by Equations \ref{eq:beta3L} and \ref{eq:gamma3L} reduce to
\begin{equation}
\beta_{2L} = \frac {3} {\sqrt{2}} \frac{e^3 \hbar^3}{E_{10}^{7/2}}\left(\frac{N}{m}\right)^{3/2} X\sqrt{1-X^4}
\label{eq:beta2L}
\end{equation}
and
\begin{equation}
\gamma_{2L} = \frac{e^4 \hbar^4 N^2}{m^2 E_{10}^5} \left(4 - 5X^4\right) ,
\label{eq:gamma2L}
\end{equation}
respectively, where the $2L$ subscript represents the system for $E_{20}\rightarrow\infty$. In this case of classically interacting molecules, there are no shifts in the energies of the eigenstates.  For unperturbed molecules, $\gamma\rightarrow \infty$ for $r^\prime \rightarrow 0$ - clearly an unphysical result when the two molecules are separated by a distance that is less than there electromagnetic diameter.  Therefore, we will restrict our discussion to the intrinsic value of molecules with the same number of electrons as two separate molecules and restrict the space to the physically-allowed region defined by $r^\prime > \sqrt[3]{2}$.

Figure \ref{fig:geffunpert} shows the semi-classical intrinsic effective second hyperpolarizability, $\gamma_{\mathrm{eff}}/\gamma_{\mathrm{max}}^\prime$, which diverges as $r\rightarrow 0$. This is a consequence of the classical model which assumes that the eigenenergies do not change even when the molecules interact with each other. Although the function diverges at zero separation, the second hyperpolarizability due to cascading never exceeds the fundamental limit when the separation distance is larger than the electromagnetic size, $r > \sqrt[3]{2 \alpha_{\mathrm{max}}}$.

\section{Real systems}
\label{sec:realsystems}

In this paper and in the companion article, we have considered cascading only between molecules with fixed orientation. The degrees of freedom offered by cascading include the separation between molecules and their orientations. The average distance between molecules, the parameter studied by Baev {\em et al}, can be easily controlled by adjusting the density of molecules in solution. The additional degree of freedom originating in the orientations of the molecules can be controlled by the application of an external electric field, which will also affect the local pair orientational correlation function due to induced dipole moments. Alternatively, local molecular forces, as are found in liquid crystals, could also have alignment-mediated enhancements. These real systems will lead to a cascading contribution that is {\em lower} than for an ideal fixed configuration.

To understand how an externally applied electric field can be used to control cascading, we develop a thermodynamic model of poling that includes the effects of the local reaction fields as treated with the self-consistent field model. In a gas, the molecules will relax when the external field is turned off, but in a poled polymer,\cite{singe87.01} the alignment can be made permanent.

An external electric field applied to an ensemble of molecules in thermal equilibrium will result in a distribution of orientations centered along the field direction.  External stress can be used to further control the orientations, and therefore, the tensor properties of the nonlinear optical response.\cite{kuzyk89.03} Here, we will consider the effects of an applied electric field only, and imagine that the orientational order is frozen in place as it would be in the case of a dye-doped polymer.

We can understand the physics of the problem as follows.  The external field tends to align the molecules.  However, the induced dipole moment of each molecule adds to the field at the site of the other molecule, where the local field at each molecule must be solved self-consistently.  Thus, the orientational order of each molecule will depend on both the applied electric field as well as the distance between the molecules.  We label this as the correlated case.

If interactions between the induced dipole moments of each pair of molecules are not taken into account, as is the case in early models of poling, cascading effects will be enhanced, but to a smaller degree. The difference between the two cases is a measure of the local correlation strength. In each case, we assume that the orientational order has been frozen in place by the poling field after it has been turned off, and that the subsequently-applied optical field, which we approximate in the zero frequency limit to describe the off-resonance response, probes this orientational distribution.

It is worth mentioning the differences between our approach and that of Baev {\em et al}. We are treating the end-to-end case, which our self consistent field calculations show to yield the largest response. In contrast, Baev {\em et al} use the side-by-side geometry. Additionally, we take into account the distribution of orientations using a thermodynamic model rather than using a static orientation.

Consider a pair of para-Nitroaniline (pNA) molecules in an ensemble, which are free to rotate in three dimensions. The angular probability density for the $i$th molecule, $P_i$, at an azimuthal angle, $\phi_i$, and polar angle, $\theta_i$, from Boltzmann statistics is
\begin{widetext}
\begin{equation}
P_i \left(\phi_i,\theta_i,\phi_j,\theta_j\right) = \displaystyle\frac{\displaystyle e^{\int_{0}^{\mathbf{E}}\mathbf{p}_i\cdot\left(d\mathbf{E}_{a}^{\prime}+d\mathbf{E}_{j}^{\prime}\right)/kT}}{\int_{0}^{\pi} \int_{0}^{2\pi} \int_{0}^{\pi} \int_{0}^{2\pi} d\phi_{i} d\theta_{i} d\phi_{j} d\theta_{j}\, \sin\theta_{i} \sin\theta_{j} e^{\int_{0}^{\mathbf{E}}\mathbf{p}_i\cdot\left(d\mathbf{E}_{a}^{\prime}+d\mathbf{E}_{j}^{\prime}\right)/kT}} ,
\label{eq:boltzmannpE}
\end{equation}
\end{widetext}
where $k$ is the Boltzmann constant, $T$ is the temperature, $\mathbf{E}_j$ is the electric field due to the $j$th molecule, $\theta$ is the polar angle measured from the $z$-axis, and $\phi$ is the azimuthal angle. Note that for two molecules, $1$ and $2$, there are only two possible combinations, i.e. $(i,j)=(1,2)$ and $(i,j)=(2,1)$.

The exponents in Equation \ref{eq:boltzmannpE} can be expanded in a power series to fourth order in the applied electric field by using the molecule's {\em effective molecular susceptibilities},
\begin{widetext}
\begin{eqnarray}
& & \int_{0}^{\mathbf{E}} \mathbf{p}_i\cdot\left(d\mathbf{E}_a+d\mathbf{E}_j\right) = \int_{0}^{\mathbf{E}_a} \mathbf{p}_i^{\mathrm{eff}} \cdot d\mathbf{E}_a = \nonumber \\
& & \mu E_{a} \cos\theta_i + \frac{1}{2}\alpha_{\mathrm{eff},i}\left(\phi_{ij},\theta_i,\theta_j\right)E_{a}^{2} \cos \theta_i + \frac{1}{3}\beta_{\mathrm{eff},i}\left(\phi_{ij},\theta_i,\theta_j\right)E_{a}^{3} \cos \theta_i + \frac{1}{4}\gamma_{\mathrm{eff},i}\left(\phi_{ij},\theta_i,\theta_j\right)E_{a}^{4} \cos \theta_i ,
\label{eq:rewriteprobabilityterm}
\end{eqnarray}
\end{widetext}
where the cosines in Equation \ref{eq:rewriteprobabilityterm} orinate from $\mathbf{E}_a\cdot\hat{f}_i = E_a \cos \theta_i$, $\mathbf{p}_i^{\mathrm{eff}}$ is the effective induced dipole moment that takes into account interactions between molecules using self-constant fields, and the permanent dipole moment, $\mu = 6.2\times 10^{-18}\,$erg$^{1/2}$cm$^{2}$ \cite{chen89.02} is now included to accurately describe the correlation fields for pNA. To find the contribution of microscopic cascading, the denominators in the $i$th molecule's effective susceptibilities, which act as the local field factors due to the polarizability of the other molecule, are expanded in a series and only the first term is kept.

The orientation-averaged contribution to the $i$th molecule's effective second hyperpolarizability, $\left\langle \gamma_{\mathrm{eff},i} \right\rangle$, is then given by
\begin{widetext}
\begin{equation}\label{eq:angularAveraging}
\left\langle \gamma_{\mathrm{eff},i} \right\rangle = \int_{0}^{\pi} \int_{0}^{2\pi} \int_{0}^{\pi} \int_{0}^{2\pi} d\phi_{i} d\theta_{i} d\phi_{j} d\theta_{j}\, \sin\theta_{i} \sin\theta_{j} \cos \theta_i P_i \left(\phi_{ij}, \theta_i, \theta_j \right) \gamma_{\mathrm{eff},i} \left(\phi_{ij}, \theta_i, \theta_j \right) .
\end{equation}
\end{widetext}
Furthermore, $\left\langle \gamma_{\mathrm{eff},1} \right\rangle = \left\langle \gamma_{\mathrm{eff},2} \right\rangle$ due to the fact that the molecules are indistinguishable. Thus, the orientationally-averaged effective second hyperpolarizability due to the two-molecule system is $\left\langle \gamma_{\mathrm{eff}} \right\rangle = 2\left\langle \gamma_{\mathrm{eff},i} \right\rangle$.

Para-Nitroaniline (pNA) is a donor-acceptor molecule that has been studied in great detail, and it is often used to test theoretical models.\cite{datta03.01,szost07.01,chen89.02} It can be approximated as a one-dimensional molecule due to its strong push-pull behavior from the opposing ends of the benzene group. This molecule has previously been the subject of density functional theory calculations, where the second hyperpolarizability was reported to be enhanced due to microscopic cascading in the side-by-side geometry.\cite{baev10.01} Here, we test the case of the effective second hyperpolarizability in the end-to-end geometric configuration in the presence of the poling field.

The three-level ansatz will give accurate results for molecules with susceptibilities near the limit.\cite{kuzyk09.01} Since pNA is far from the limit ($\beta$ and $\gamma$ are each about 2 orders of magnitude below the fundamental limit), we will use the measured values of the linear and nonlinear molecular susceptibilities instead of the calculated ones; but, will ignore the changes in the transition energies that arise from molecular interactions.  Figures \ref{fig:alignedcutoff} and \ref{fig:geffunpert} show the regions where the classical model of cascading is a fair approximation, that is, $r^\prime > \sqrt[3]{2}$. The susceptibilities we use here are, $\alpha = 1.7\times 10^{-23}\,$cm$^{3}$, $\beta = 9.2\times 10^{-30}\,$erg$^{-1/2}$cm$^{4}$, and $\gamma = 1.5\times 10^{-35}\,$erg$^{-1}$cm$^{5}$ -- values that were reported by Cheng \textit{et al}.\cite{chen89.02}

Equations \ref{eq:fullalpha}-\ref{eq:fullgamma} in the appendix show the effective polarizability, hyperpolarizability and second hyperpolarizability of molecule $i$ in the presence of molecule $j$ for two arbitrary but fixed orientations, solved self-consistently in analogy to the simple end-to-end case given by the derivation leading to Equations \ref{eq:align1Dalpha} through \ref{eq:align1Dgamma}. The orientational averages of these results are calculated according to Equation \ref{eq:angularAveraging}. These calculations, which are far too complex to solve analytically, were evaluated numerically using Matlab.\circledR

First, we calculate the cascading contributions for the uncorrelated angular distribution for the tensor components along the applied field. For a molecule that makes an angle $\theta$ with the field, the effective polarizability and (second) hyperpolarizabilities along the field are given by,
\begin{eqnarray}
\alpha_{\mathrm{eff},i}^{\mathrm{uncorr}} = \alpha \cos \theta , \label{eq:uncorralpha} \\
\beta_{\mathrm{eff},i}^{\mathrm{uncorr}} = \beta \cos^2 \theta , \label{eq:uncorrbeta}
\end{eqnarray}
and
\begin{equation}
\gamma_{\mathrm{eff},i}^{\mathrm{uncorr}} = \gamma\cos^3 \theta + 4 \frac{\beta^2 \cos^4 \theta}{r^3} .
\label{eq:uncorrgamma}
\end{equation}
The subscript $i$ has been dropped from the angle because both molecules are identical and have the same orientational distribution function due to the applied electric field.  Then, the orientational average of Equation \ref{eq:uncorrgamma} is determined using the orientational distribution function that includes only the effects of the applied poling electric field.   The correlated case is too complex to express, and is calculated using the self-consistent field calculation to determine the orientational distribution function from which the orientational average of the effective second hyperpolarizability is calculated.

\begin{figure}[t!]
\centering\includegraphics{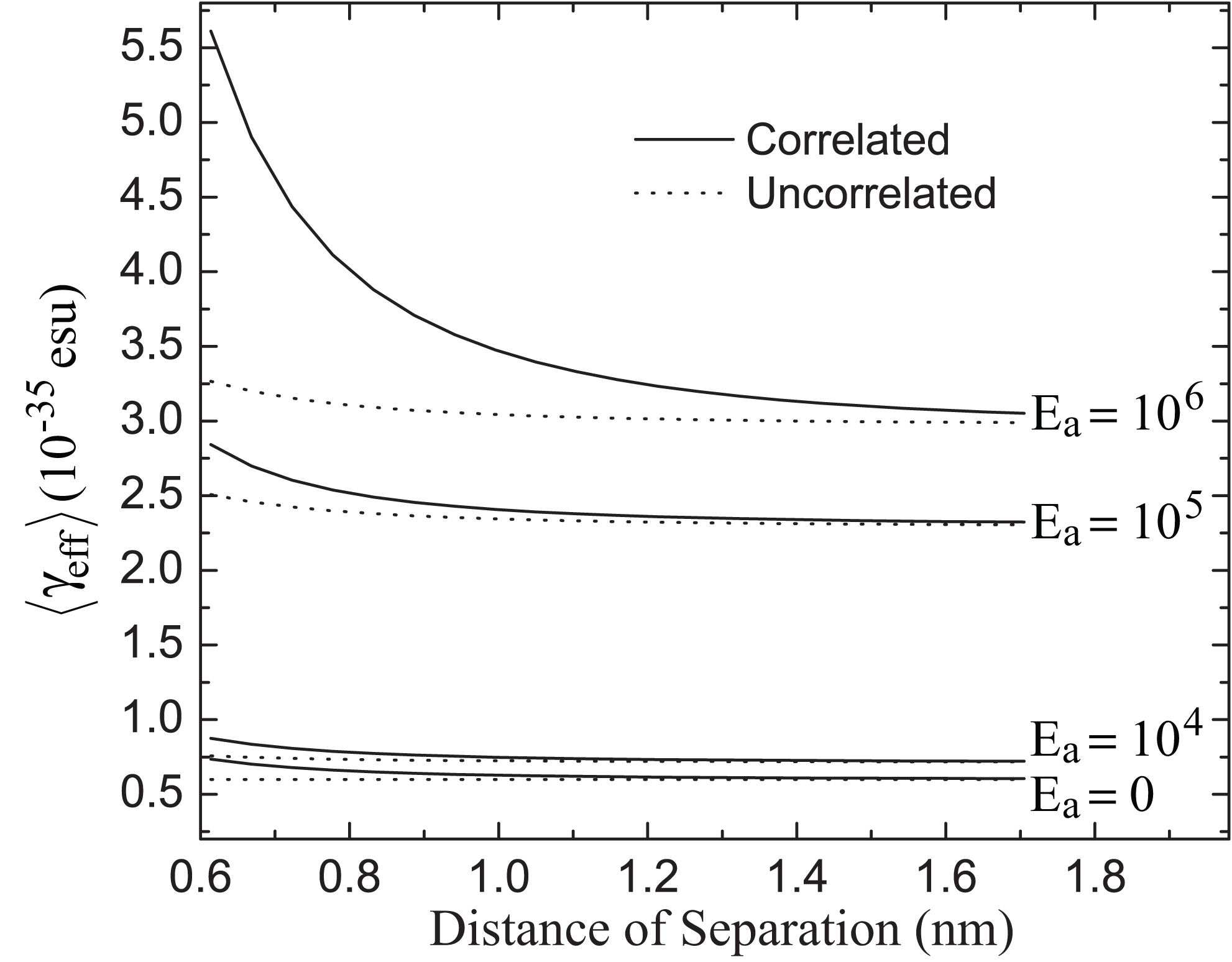}
\caption{The effective second hyperpolarizability of two interacting pNA molecules at a temperature of $297\,$K where the separation vector between the two molecules is along the direction of the applied field. The molecules are freely rotating, where the alignment is a result of the applied electric field in units of StatV/cm  The solid curves include interactions between polarized molecules and the dotted curves include only the effects of the applied electric field.}
\label{fig:pNAaligned}
\end{figure}
Figure \ref{fig:pNAaligned} shows the orientational-averaged effective second hyperpolarizability, $\left\langle \gamma_{\mathrm{eff}} \right\rangle$, for the correlated (solid curves) and uncorrelated cases (dotted curves) as a function of the distance of separation, $r$, for several electric field strengths. At large separation and zero applied field, the orientational average of the effective second hyperpolarizability is $\left\langle \gamma_{eff} \right\rangle = 2 \gamma_{xxxx} / 5$, as expected.

The applied poling field at a field strength of $E_a = 10 ^6$ at $6 \AA$ separation (the approximate size of the molecule) enhances the effective response, which includes cascading, by appoximately a factor of four while correlations due to induced dipole interactions leads to another factor of about two. Thus poling correlations add constructively to the orientational ordering effects of the field, leading to approximately an order-of magnitude enhancement.  Note this this is far short of the largest possible enhancements for perfectly aligned molecules due to thermal disordering effects.

In this example, while the net enhancement effect of the intrinsic hyperpolarizability is about an order of magnitude relative to the sum of the nonlinearities of two individual molecules, the largest possible nonlinear response when the two molecules are combined together into a single molecule would give a gain of a factor of 2.  Thus, the enhancement effect due to cascading yields an increase in the intrinsic hyperpolarizability by about a factor of 5.  Thus, while cascading may not break the limit of the nonlinear-optical response, it can lead to an increase in the nonlinear response over the best available molecules for applied fields that are at the dielectric breakdown limits for the best materials.

\begin{figure}[t!]
\centering\includegraphics{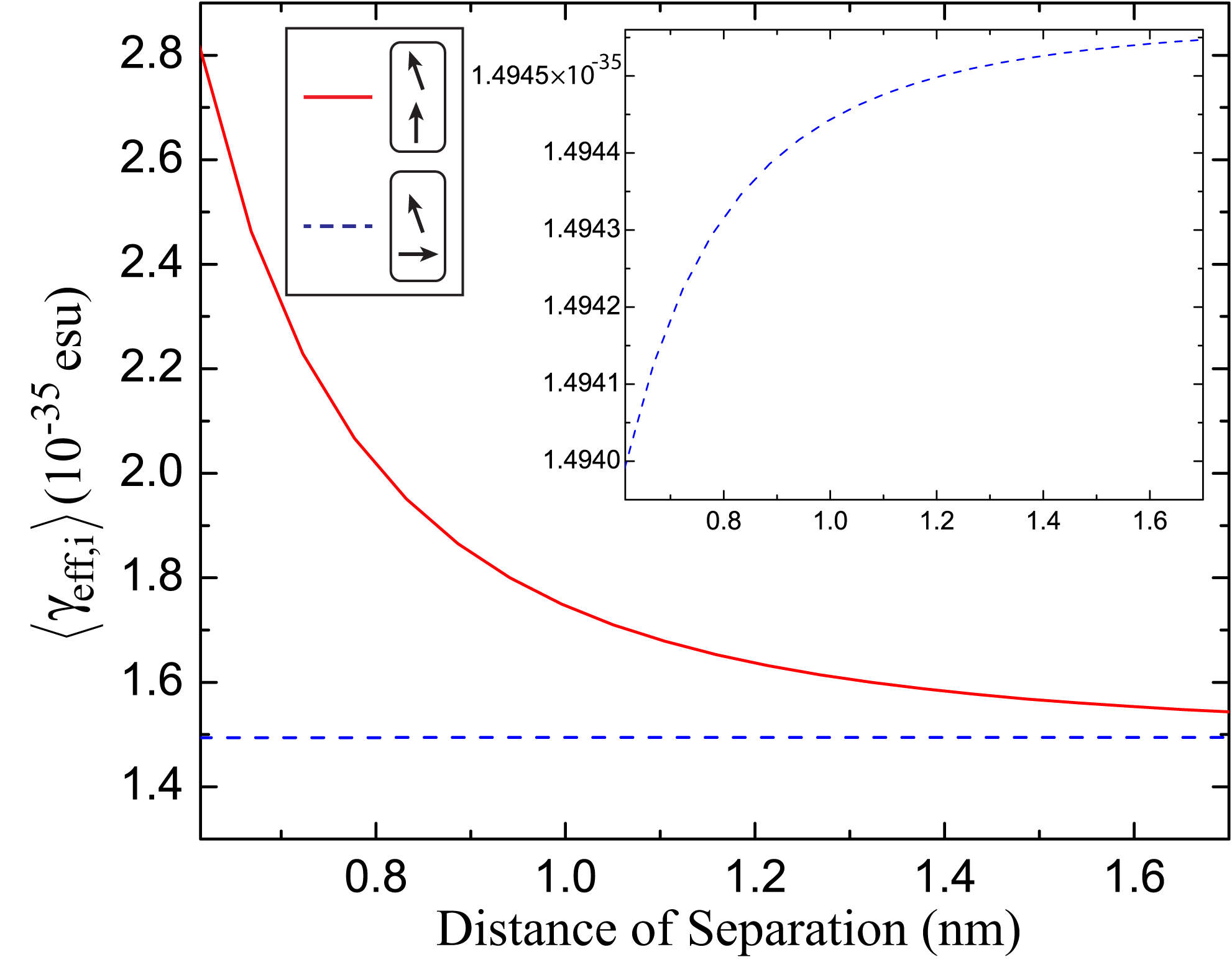}
\caption{The contribution to the effective second hyperpolarizability of one freely rotating molecule that is coupled to a second molecule. The second molecule is held in both the direction perpendicular and parallel to the applied field while the other is free to move.  The temperature is $297\,$K and the poling field is $10^6\,$StatV/cm.}
\label{fig:perpparaPNA}
\end{figure}
The coupling strength between two molecules depends on the distance between them, and is a key parameter for cascading. The effect of the strength of interaction on the orientation can be isolated by allowing one molecule to rotate freely and the other to be fixed with its long axis either perpendicular or parallel to the applied field. Figure \ref{fig:perpparaPNA} shows $\left\langle \gamma_{\mathrm{eff},i} \right\rangle$ for these two orientations of the fixed molecule. The inset, which shows a magnified view of the effective second hyperpolarizability of the perpendicular orientation, illustrates the weakness of the intermolecular interaction from orientational averaging in the presence of a strong applied fields when the fixed molecule is held perpendicular to it.

These results give insights into cascading and how extra degrees of freedom can be used to design molecular systems with a larger intrinsic nonlinear-optical response. It is clear that cascading of lower-order nonlinear molecules with additional control of orientational order, as would be the case for liquid crystals or mechanically-stressed polymers, would further increase the microscopic cascading contribution due to the control of additional order parameters.

\section{Approximations}

In this section, we briefly discuss the approximations used in our calculations, and how they may limit the validity of the results.

{\bf Off-resonance response.}  All of the calculations consider the zero-frequency limit of the electronic response. In real experiments, the optical field frequencies can be tuned to be smaller than transition frequencies within the molecule, but there will always be some resonance enhancement.  These can be partially taken into account by using the two level dispersion model by Oudar and Chemla.\cite{oudar77.03}

{\bf Dipole approximation:} Our calculations assume that the induced dipole moment can be expanded in a series of the electric field of the form $p = \alpha E + \beta E^2 + ...$.  Many nonlinear-optical processes originate under the assumption that the series converges.  This implicitly assumes that the fields must be small, that is, $\gamma E / \beta \ll 1$, etc. to all orders.  Thus, the molecules must not get close enough together to violate this condition.

{\bf Finite physical size:} It is assumed that the separation between the molecules must be larger than their size, otherwise the two molecules will merge together to form a new one.  The electromagnetic size of a molecule is determined from the polarizability $\alpha$ to be $r > (2 \alpha) ^{3/2}$.

{\bf Three-level ansatz:} The three-level ansatz states that when a molecule has a nonlinear response near the fundamental limit, it can be modeled as a three-state system using a sum-rule-restricted reduced-parameter space.  While this has not been proven rigourously, it is an empirical observation that holds for a wide rage of systems.\cite{kuzyk09.01} Thus, results in the regime of small nonlinear response are inaccurate; but, near the limit -- the focus of this paper, the calculations are accurate.

{\bf Magnetic interactions:} We assume that the electromagnetic vector potential can be ignored, so that that the position matrix is real.  As a consequence, $x_{ij} = x_{ji}$.  This assumption does not change the generality of the results.\cite{watkins09.01}

{\bf Perturbation approximation:} The energy shifts and change in dipole matrix elements are calculated using perturbation theory; so, the molecules must be far enough apart to limit interaction energies to be smaller than the eigenenergies of each molecule.  Details of why this approximation holds at all distances is described in the companion article.\cite{dawson11.01a}

{\bf pNA} In the calculation of the self-consistent field interactions between two pNA molecules, the energy shifts and dipole matrix changes are neglected.  This approximation will not hold when the separation between molecules gets small, so the calculated results will not be accurate in the limit of small separations.

{\bf 1D molecules:} This approximation is commonly used for push-pull molecules and does not lead to large deviations from the non-idealized case. The uncertainty introduced by this approximation for a system with two nonzero tensor components is about $\xi_x / \sqrt{\xi_x^2 + \xi_y^2}$. For example, if the next larger tensor component is 1/3 of the largest one, this will introduce an error of 5\%.

See the companion article for a more detailed discussion of the physics behind these approximations.\cite{dawson11.01a}

\section{Conclusion}
\label{sec:discusscasc}

In this work, we have investigated cascading as a method for enhancing the second hyperpolarizability. The fundamental limit has been used to define the scale-invariant intrinsic hyperpolarizabilities, which enables us to assess the effectiveness of cascading as a means of producing nonlinear susceptibilities larger than what is possible for a single molecule.

In the case of two one-dimensional molecules in an end-to-end configuration aligned with the direction of the applied electric field, the cascading contribution to the second hyperpolarizability was shown to be bounded by the fundamental limit provided that $r^3 > 2\alpha_{\mathrm{max}}$, a requirement that the two molecules do not overlap. Additionally, this condition ensures that the fundamental basis of nonlinear optics, i.e. that the polarization can be expressed as a power series of the electric field, is valid by ensuring convergence.

Our full theory uses the self-consistent field approach to derive the effective nonlinear susceptibilities of the two-molecule system, and includes the effects of van der Waals interactions on the energies and transition moments of each molecule.  This full quantum treatment shows that the two-molecule system is bounded by the same fundamental limit as one molecule, the same result as we get in the other extreme case of two side-by-side molecules.\cite{dawson11.01a}  Thus, we conclude that combining two molecules into one or allowing two molecules to interact with cascading leads to the same optimized response.

The bottom line is that cascading is not a loophole that allows the fundamental limits to be exceeded. However, cascading provides an additional degree of freedom for designing materials, and therefore may provide an alternative approach for improving the nonlinear response over what is possible through synthetic means alone.  As we show for the model system pNA, cascading can yield an enhancement of the intrinsic second hyperpolarizability, i.e. the response can get closer to the limit.  It may be difficult to manipulate nonlinear responses with a single molecule design of the same size as the two-molecule system that interacts with cascading.

{\bf Acknowledgements}  We thank the National Science Foundation (ECCS-0756936) and the Air Force Office of Scientific Research (FA9550-10-1-0286) for generously supporting this work.

\begin{widetext}

\section{Appendix: Effective molecular susceptibilities for two freely-rotating bodies}
\label{app:freerotmol}

For two molecules of arbitrary fixed orientations and separation vector along the applied electric field, the effective polarizability of molecule $i$ is derived from Equations \ref{eq:molsus3rdorder}-\ref{eq:kn}, and yields
\begin{eqnarray}
\alpha_{\mathrm{eff},i} = r^3 \alpha\frac{\left(r^3 + \alpha \cos\left(2\theta_j\right) + \alpha \right)\cos\theta_i - \alpha \cos\phi_{ij} \sin\theta_i \sin\theta_j \cos\theta_j }{r^6+\alpha^2 \left[\cos\phi_{ij} \left( \sin\left(2 \theta _i \right) \sin\left(2\theta_j \right) -\cos \phi_{ij} \sin^2 \theta_i \sin^2 \theta_j \right) -4 \cos^2 \theta_i \cos^2 \theta_j \right]} ,
\label{eq:fullalpha}
\end{eqnarray}
where the local field factor is automatically and naturally taken into account through the self-consistent field calculation. The effective polarizability of the two molecules together is $\alpha_{\mathrm{eff}} = \alpha_{\mathrm{eff},1} + \alpha_{\mathrm{eff},2}$. The next term given by Equation \ref{eq:kn} is the $i$th molecule's contribution to the effective hyperpolarizability, which gives
\begin{eqnarray}
\beta_{\mathrm{eff},i} &=& r^9 \beta \left[4 r^3 \alpha \cos^3 \theta_j \cos \theta_i - 4r^3 \alpha  \cos\theta_j \cos^3 \theta_i + 2 \alpha \cos^2 \theta_j \left(4r^3 \cos^2 \theta_i + 4 \alpha \cos^4 \theta_i - r^3 \cos\phi_{ij} \sin\theta_j \sin\theta_i \right) \right. \nonumber \\
&+& \left. 2\cos^2 \theta_i \left(r^6+\alpha \cos\phi_{ij} \sin\theta_i \left(-2 \alpha  \cos\theta_i \sin\left(2\theta_j\right) + \sin\theta_j \left(r^3 + \alpha \cos\phi_{ij} \sin\theta_j \sin\theta_i\right)\right)\right) \right. \nonumber \\
&-& \left. r^3 \alpha \cos\phi_{ij} \sin\left(2\theta_j\right) \sin\left(2\theta_i\right)\right] \nonumber \\
&/& \left[2 \left(r^3+2 \alpha  \cos\theta_j \cos\theta_i - \alpha \cos\phi_{ij} \sin\theta_j \sin\theta_i\right)^2 \left(r^3 - 2\alpha \cos\theta_j \cos\theta_i + \alpha \cos\phi_{ij} \sin\theta_j \sin\theta_i\right)^3\right] .
\label{eq:fullbeta}
\end{eqnarray}

The effective second hyperpolarizability is found in the same way. The $i$th molecule's contribution to the effective second hyperpolarizability of a system of two fixed molecules in the end-to-end configuration is given by,
\begin{eqnarray}
\gamma_{\mathrm{eff},i} &=& \left\{ \left[ r^6 \alpha \beta^2 \left( -2 \cos\theta_i \cos\theta_j + \cos\phi_{ij} \sin\theta_i \sin\theta_j\right)^2 \left(\cos\theta_j\left(r^3 + \alpha \cos\left( 2\theta_i \right) + \alpha \right) - \alpha \cos\left(\theta_i \right)\cos\phi_{ij}\sin\theta_i\sin\theta_j \right) \right. \right. \nonumber \\
&\times& \left(4 r^3 \alpha \cos\theta_i \cos^3 \theta_j - 4 r^3 \alpha \cos^3 \theta_i \cos\theta_j - 4\alpha^2 \cos \phi_{ij} \cos^3 \theta_i \sin\theta_i \sin\left(2\theta_j \right) \right. \nonumber \\
&+& 2\alpha\cos^2 \theta_j \left(4 r^3\cos^2\theta_i + 4\alpha\cos^4 \theta_i - r^3 \cos \phi_{ij} \sin\theta_i \sin\theta_j \right) \nonumber \\
&+& \left. \left. 2 \cos^2 \theta_i \left(r^6 + r^3\alpha \cos\phi_{ij}\sin\theta_i \sin\theta_j + \alpha^2 \cos^2 \phi_{ij} \sin^2 \theta_i \sin^2 \theta_j \right) - r^3 \alpha \sin \left(2\theta_i\right)\sin\left(2\theta_j \right) \cos\phi_{ij} \right) \right] \nonumber \\
&/& \left[ \left( r^3 + 2\alpha \cos\theta_j \cos\theta_i - \alpha \cos\phi_{ij} \sin\theta_i \sin\theta_j\right)^3 \left( r^3 - 2\alpha\cos\theta_i \cos\theta_j + \alpha \cos\phi_{ij} \sin\theta_i \sin\theta_j \right)^4 \right] \nonumber \\
&+& \frac {r^9 \gamma \left(\cos\theta_i \left( r^3 + \alpha \cos\left(2\theta_j \right) + \alpha \right) - \alpha \cos\phi_{ij} \sin\theta_i \sin\theta_j \cos\theta_j \right)^3} {\left (r^6 - 4\alpha^2 \cos^2 \theta_i \cos^2 \theta_j - \alpha^2 \cos^2 \phi_{ij}\sin^2 \theta_i \sin^2 \theta_j + \alpha^2 \cos\phi_{ij}\sin\left(2\theta_i \right) \sin\left( 2\theta_j \right) \right)^3} \nonumber \\
&+& \frac {r^6\alpha\gamma \left(2\cos\theta_i \cos\theta_j - \cos\phi_{ij}\sin\theta_i \sin\theta_j \right) \left(\cos\theta_j \left(r^3 + \alpha\cos\left(2\theta_i \right) + \alpha \right) - \alpha \cos\phi_{ij} \cos\theta_i \sin\theta_i \sin\theta_j \right)^3} {\left (r^6 - 4\alpha^2 \cos^2 \theta_i \cos^2 \theta_j - \alpha^2 \cos^2 \phi_{ij}\sin^2 \theta_i \sin^2 \theta_j + \alpha^2 \cos\phi_{ij}\sin\left(2\theta_i \right) \sin\left( 2\theta_j \right) \right)^3} \nonumber \\
&+& \left[ r^9 \beta^2 \left(2\cos\theta_i \cos\theta_j - \cos\phi_{ij} \sin\theta_i \sin\theta_j \right) \left( 2 \cos\theta_i \left( r^3 + \alpha \cos\left(2\theta_j \right) + \alpha \right)  - \alpha\cos\phi_{ij} \sin \left( 2\theta_j \right) \sin\theta_i \right) \right. \nonumber \\
&\times& \left( 16 \alpha^2 \cos^4 \theta_j \cos^2 \left(\theta_i\right) + 8 r^3 \alpha \cos^3 \theta_i \cos\theta_j + 4 r^3 \cos^2 \theta_j \left(r^3 + 4\alpha \cos^2 \theta_i + \alpha\cos\phi_{ij} \sin\theta_i \sin\theta_j \right) \right. \nonumber \\
&-& 8\alpha\cos^3 \theta_j \left(r^3 \cos\theta_i + \alpha \cos\phi_{ij} \sin\left(2\theta_i \right) \sin\theta_j \right) \nonumber \\
&-& \left. \left. \alpha \cos\phi_{ij}\left( 4 r^3 \cos^2 \theta_i \sin\theta_i \sin\theta_j + \sin\left( 2\theta_j \right) \left( 2 r^3 \sin\left( 2\theta_i \right) - \alpha \cos\phi_{ij} \sin\left(2\theta_j \right) \sin^2 \theta_i \right) \right) \right) \right] \nonumber \\
&/& \left. \left[ 4\left( r^3 + 2\alpha \cos\theta_i \cos\theta_j - \alpha \cos\phi_{ij} \sin\theta_i \sin\theta_j \right)^3 \left(r^3 - 2\alpha \cos\theta_i \cos\theta_j + \alpha \cos\phi_{ij} \sin\theta_i \sin\theta_j \right)^4 \right] \right\} \nonumber \\
&/& \left[ 1 - \frac{\alpha^2}{r^6} \left( -2\cos\theta_i \cos\theta_j + \cos\phi_{ij} \sin\theta_i \sin\theta_j \right)^2 \right] .
\label{eq:fullgamma}
\end{eqnarray}

These results are used extensively is the section on real molecules. Equations \ref{eq:fullalpha} through \ref{eq:fullgamma}, which describe the effective molecular susceptibilities of two interacting molecules, are exact for a uniform applied electric field and a nonlinear-optical response that can be represented as a power series in the electric field vector.

\end{widetext}

\end{document}